# Low wavenumber Raman spectroscopy of highly crystalline MoSe$_2$ grown by chemical vapor deposition

**Maria O'Brien**[‡1,2], **Niall McEvoy**[‡2], **Damien Hanlon**[2,3], **Kangho Lee**[1,2], **Riley Gatensby**[1,2], **Jonathan N. Coleman**[2,3], and **Georg S. Duesberg**[*1,2]

[1] School of Chemistry, Trinity College Dublin, Dublin 2, Ireland
[2] Centre for Research on Adaptive Nanostructures and Nanodevices (CRANN) and Advanced Materials and BioEngineering Research (AMBER) Centre, Trinity College Dublin, Dublin 2, Ireland
[3] School of Physics, Trinity College Dublin, Dublin 2, Ireland
* Corresponding author: duesberg@tcd.ie
‡ These authors contributed equally to this work

**Keywords** Raman Spectroscopy, Chemical Vapor Deposition, Transition Metal Dichalcogenides, MoSe$_2$

Transition metal dichalcogenides (TMDs) have recently attracted attention due to their interesting electronic and optical properties. Fabrication of these materials in a reliable and facile method is important for future applications, as are methods to characterize material quality. Here we present the chemical vapor deposition of MoSe$_2$ monolayer and few layer crystals. These results show the practicality of using chemical vapor deposition to reliably fabricate these materials. Low frequency Raman spectra and mapping of shear and layer breathing modes of MoSe$_2$ are presented for the first time. We correlate the behavior of these modes with layer number in the materials. The usefulness of low frequency Raman mapping to probe the symmetry, quality, and monolayer presence in CVD grown 2D materials is emphasized.

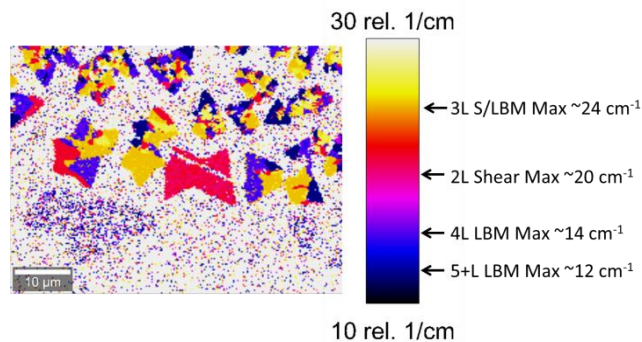

Raman map of maximum intensity position for low wavenumber shear and layer breathing modes of MoSe$_2$.

**1 Introduction** Transition metal dichalcogenides (TMDs) such as MoS$_2$, WS$_2$, MoSe$_2$ and WSe$_2$ have recently attracted significant attention from the research community due to their novel properties that make them desirable for numerous applications[1-4]. In particular, in 2010, Mak et al.[5] and Splendiani et al.[6] showed that molybdenum disulfide (MoS$_2$), which had previously been shown to exist in few layer state,[7] showed extraordinary photoluminescence in its single layer form, indicating a change in the electronic structure of MoS$_2$ with layer number. The discovery of a significant electron mobility with dielectric encapsulation[8] as well as a direct bandgap (~1.8 eV)[5, 6, 9] in monolayer MoS$_2$ and other atomically thin TMDs[10, 11] has led to the suggestion that they may be more suitable in applications at the ultimate thickness limit for electronics. More recent reports have suggested that transition metal diselenides such as MoSe$_2$ may be more suitable than transition metal disulfides for device applications due to its narrower bandgap of 1.5 eV with a similarly high mobility[11]. While techniques such as mechanical exfoliation can produce sufficient material to indicate initial suitability for these applications, reliable growth methods need to be realized for long term development.

Raman spectroscopy is a powerful and non-destructive technique for assessing the electronic and structural properties of TMDs for device application suitability. While initial reports on Raman of MoSe$_2$ and WSe$_2$ focus on the high wavenumber Raman modes in these materials, there is a noticeable lack of Raman mapping of low wavenumber modes of transition metal diselenides to show layer number, uniformity, single crystals and grain boundaries. The layer number sensitivity of low wavenumber Raman modes of these materials means they could act as a simple verification of layer number and sample quality for these materials for a variety of applications, the most pressing of which is research into the development of pristine layers into electronic devices.

Here we demonstrate the growth of MoSe$_2$ by chemical vapor deposition (CVD) in a microreactor setup, demonstrating that a previously reported synthesis method[12] is extendable across other TMD systems. We also show for the first time the mapping of low wavenumber Raman modes in MoSe$_2$, and provide a reliable method of using these modes to determine layer number of the material.

**2 Experimental** Precursor layers of MoO$_3$ were liquid phase exfoliated and dispersed onto commercially available silicon dioxide (SiO$_2$, ~290 nm thick) substrates[12, 13]. A TEM image of a typical MoO$_3$ nanosheet is shown in Figure 1(a), with an optical image of a nanosheet dispersion. The MoO$_3$ precursor substrates were placed face-up in a ceramic boat with a blank SiO$_2$/Si substrate placed face down on top of this, in order to direct growth to the top, clean substrate, as described previously[12], forming a microreactor. This is illustrated in the schematic in Figure 1(a). This was then placed in the centre of the heating zone of a quartz tube furnace, and ramped to 750 $^o$C under 150 sccm of 10% H$_2$/Ar flow. Selenium (Se) vapor was then produced by heating Se powder to ~220 $^o$C in an independently controlled upstream heating zone of the furnace, and carried downstream to the microreactor for a duration of 30 minutes after which the furnace was cooled down to room temperature. A schematic of the furnace setup is shown in Figure 1(b).

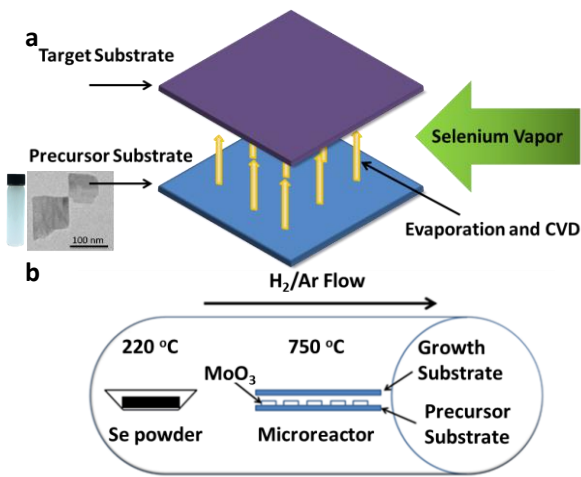

**Figure 1 -** (a) Schematic of CVD microreactor formed between the seed and target substrates, where selenium reacts with MoO$_3$ nanosheets to form MoSe$_2$ layers on the top substrate. (b) Schematic of furnace setup. Selenium powder is melted downstream and flowed through the microreactor.

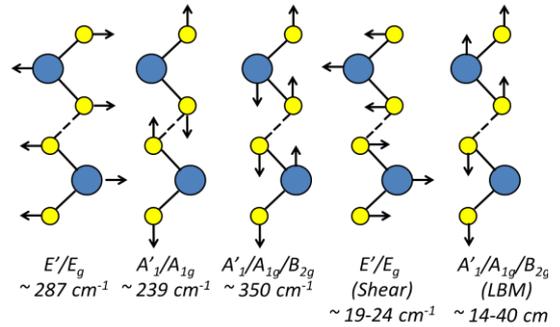

**Figure 2 -** Schematic of Raman active vibration modes in MoSe$_2$ with labels and positions below.

Raman spectroscopy was performed using a Witec alpha 300R with a 532 nm excitation laser and a laser power of < 1 mW, in order to minimize sample damage. The Witec alpha 300R was fitted with a Rayshield Coupler to detect Raman lines close to the Rayleigh line at 0 cm$^{-1}$.

**3 Results and Discussion** The Raman spectra of transition metal dichalcogenides such as MoSe$_2$ generally display two main characteristic vibration modes for in-plane and out of plane vibrations of metal and chalcogen atoms[14]. In the case of a monolayer (and other odd numbered layers) which belongs to the D$_{3h}$ symmetry group, MoSe$_2$ crystal has two main high wavenumber Raman peaks. These are $E'$ at ~ 287 cm$^{-1}$, and $A'_1$ at ~ 239 cm$^{-1}$. These arise from in-plane vibrations of Mo and Se atoms and out-of-plane vibrations of Se atoms in different directions only, respectively[15, 16]. For bilayer MoSe$_2$ (and other even numbered layers) due to the changing symmetry point group from D$_{3h}$ to D$_{3d}$, these are labelled $E_g$ and $A_{1g}$ respectively. For the transition metal disulfides, MoS$_2$ and WS$_2$, these peaks have been shown to shift in position with layer number[17, 18], allowing mono and few-layers to be identified easily and quickly. However for MoSe$_2$, these changes in frequency for different layer numbers are not as dramatic[19], with ~ 1 cm$^{-1}$ shift in $A_{1g}$ peak from 2 to 3L easily lost in instrumental spectral resolution. Therefore an alternative method for assessing MoSe$_2$ layer numbers using Raman spectroscopy is desirable.

Another Raman mode appears at ~ 350 cm$^{-1}$ for 2+ layers of MoSe$_2$, which does not appear for monolayer and bulk, known as the $A_{1g}$ (2L and other ENL), $A'_1$ (3L and other ONL) and $B_{2g}$ (bulk, which belongs to D$_{6h}$ symmetry group). To avoid confusion with the $A'_1/A_{1g}$ mode at ~ 239 cm$^{-1}$, Raman modes will be labelled with the wavenumber they appear at for the remainder of this paper, e.g. $A'_1/A_{1g}/B_{2g}$ (~ 350 cm$^{-1}$). These Raman active vibration modes are illustrated in Figure 2.

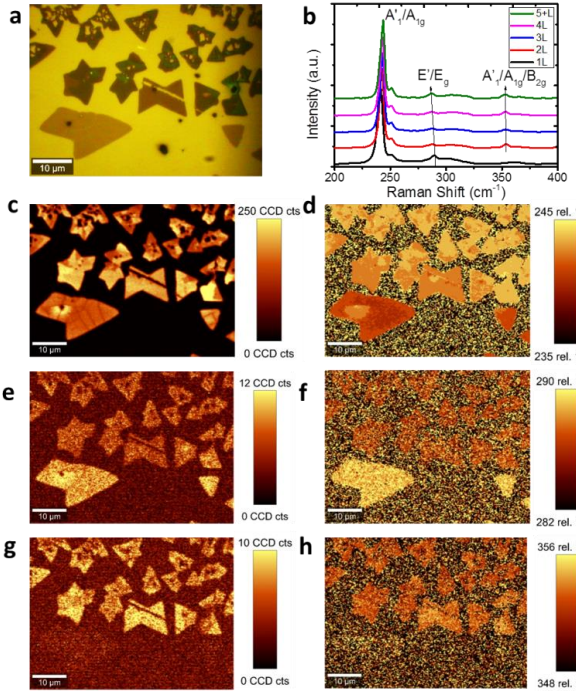

**Figure 3 –** High wavenumber Raman spectroscopy of $MoSe_2$. (a) Optical image of as-grown CVD $MoSe_2$ with varying layer numbers (b) Raman spectra of 1, 2, 3, 4, and 5+ L $MoSe_2$ (c) Map of max $A'_1/A_{1g}$ (~ 239 $cm^{-1}$) high wavenumber Raman mode (d) Position map of max $A'_1/A_{1g}$ (~ 239 $cm^{-1}$) high wavenumber Raman mode (e) Map of max $E'/E_g$ (~ 287 $cm^{-1}$) Raman mode (f) Position map of max $E'/E_g$ (~ 287 $cm^{-1}$) Raman mode (g) Map of max $A'_1/A_{1g}/B_{2g}$ (~ 350 $cm^{-1}$) Raman mode (h) Position map of max $A'_1/A_{1g}/B_{2g}$ (~ 350 $cm^{-1}$) Raman mode.

In Figure 3(a), a typical sample of as-grown $MoSe_2$ single crystals is shown. Optically, this can be identified as monolayer and few-layer $MoSe_2$ single crystals by the characteristic triangular shape[2, 11] of CVD grown TMDs. Figure 3(b) shows spectra of 1 to 5+ L $MoSe_2$ extracted from different areas in Figure 3(a), in agreement with spectra previously shown in the literature[19, 20]. A Raman map of $A'_1/A_{1g}$ (~ 239 $cm^{-1}$) intensity maximum is shown in Figure 3(c), with the corresponding maximum position in Figure 3(d). It is clear from these images that the $A'_1/A_{1g}$ (~ 239 $cm^{-1}$) intensity and position does not change dramatically as layer number increases from monolayer to bulk. Little to no variation is seen over the area mapped, bar from a slight intensity decrease in the presence of grain boundaries. A Raman map of $E'/E_g$ (~ 287 $cm^{-1}$) intensity maximum is shown in Figure 3(e), with the corresponding maximum position in Figure 3(f). This $E'/E_g$ (~ 287 $cm^{-1}$) intensity maximum and position changes significantly from monolayer to multilayer, but shows no significant change between 2, 3, 4 and 5 layers. It is more interesting to consider the intensity maximum and position maps of $A'_1/A_{1g}/B_{2g}$ (~ 350 $cm^{-1}$). As this mode does not appear for monolayer $MoSe_2$, its absence in areas serves as a direct confirmation of the monolayer nature of those areas. However, similar to $E'/E_g$ (~ 287 $cm^{-1}$), it does not shift in intensity or position for 2+ layers, which leads to difficulties in the determination of exact layer number from high wavenumber Raman spectroscopy alone.

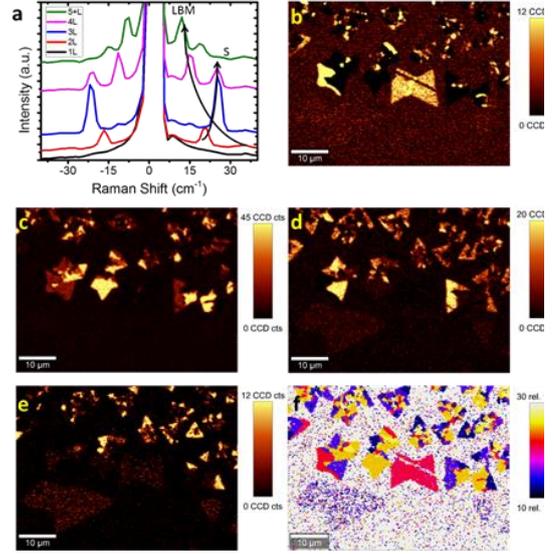

**Figure 4 -** (a) Low wavenumber Raman spectra of shear and layer breathing modes of 1, 2, 3 and 4L $MoSe_2$ (b) Map of max shear mode for 2L $MoSe_2$ at ~20 $cm^{-1}$ (c) Map of max shear/LBM mode for 3L $MoSe_2$ at ~24 $cm^{-1}$ (d) Map of max LBM mode for 4L $MoSe_2$ at ~16 $cm^{-1}$ (e) Map of max LBM mode for 5L $MoSe_2$ at ~12 $cm^{-1}$ (f) Position map of maxima of low wavenumber Raman modes, allowing fast identification of layer number.

We therefore move our attention to the low wavenumber Raman modes in these materials. These are known as the shear and layer breathing modes (LBM), and are illustrated schematically with their respective labels in Figure 2. These modes have previously been shown to be direct probes of layer number in other layered materials such as $MoS_2$[21]. Figure 4(a) shows spectra of 1 to 5+ L $MoSe_2$ which has been extracted from different areas in Figure 3(a), in agreement with spectra previously shown in the literature[19, 20]. A Raman map of the 2L shear mode $E_g$ (~ 20 $cm^{-1}$) intensity maximum is shown in Figure 4(c), with the corresponding maximum position in Figure 4(d). The enhanced intensity in certain areas demonstrates the presence of solely 2L $MoSe_2$. This can also be shown to be the case for maximum intensity maps of 3L Shear/LBM peak overlap (~ 25 $cm^{-1}$), 4L LBM (~ 15 $cm^{-1}$) and 5+ L LBM (~ 12 $cm^{-1}$) shown in Figure 4 (c), (d) and (e) respectively. The position of maximum intensity of the low wavenumber modes can also be mapped, as shown in Figure 4(f), which allows the direct determination of each layer number from a single extracted Raman image. By correlating the position of each peak of maximum intensity for each layer number, it is trivial to determine how many layers are present in each area. This is possible due to the nature of the low wavenumber modes between 10 and 30 cm$^{-1}$. Each number of layers (1 to 5+) has a peak of maximum intensity in this region, where the position of each maximum does not overlap with the position of maximum intensity of a different layer number. The low wavenumber peak positions determined here are in agreement with previously observed low wavenumber modes in mechanically exfoliated MoSe$_2$[22].

This Raman analysis reveals that micrometer sized domains of high quality crystalline MoSe$_2$ have been produced, with a variety of layer numbers that are advantageous for initial studies on optical and electronic properties. It is envisaged that further refinement of the growth procedure will lead to controlled synthesis of large area MoSe$_2$ thin films for device fabrication and further studies.

In conclusion, we have presented reliable method towards the synthesis of MoSe$_2$, using CVD in a microreactor set-up. The method yields single crystals of mono- and few layered MoSe$_2$ as revealed by detailed Raman analysis. In particular, the determination of layer number of CVD grown MoSe$_2$ by low wavenumber position mapping is introduced, which is crucial as higher frequency modes do not allow to easily extract this data.

**Acknowledgements** This work is supported by the SFI under Contract No. 12/RC/2278 and PI_10/IN.1/I3030. M.O. acknowledges an Irish Research Council scholarship via the Enterprise Partnership Scheme, Project 201517, Award 12508. NME acknowledges SFI for 14/TIDA/2329. D.H. acknowledges ERC grant SEMANTICS.

### References

1. A. K. Geim and I. V. Grigorieva, *Nature*, 2013, **499**, 419-425.
2. A. M. van der Zande, P. Y. Huang, D. A. Chenet, T. C. Berkelbach, Y. You, G.-H. Lee, T. F. Heinz, D. R. Reichman, D. A. Muller and J. C. Hone, *Nature materials*, 2013, **12**, 554-561.
3. V. Nicolosi, M. Chhowalla, M. G. Kanatzidis, M. S. Strano and J. N. Coleman, *Science*, 2013, **340**.
4. J. N. Coleman, M. Lotya, A. O'Neill, S. D. Bergin, P. J. King, U. Khan, K. Young, A. Gaucher, S. De, R. J. Smith, I. V. Shvets, S. K. Arora, G. Stanton, H.-Y. Kim, K. Lee, G. T. Kim, G. S. Duesberg, T. Hallam, J. J. Boland, J. J. Wang, J. F. Donegan, J. C. Grunlan, G. Moriarty, A. Shmeliov, R. J. Nicholls, J. M. Perkins, E. M. Grieveson, K. Theuwissen, D. W. McComb, P. D. Nellist and V. Nicolosi, *Science*, 2011, **331**, 568-571.
5. K. F. Mak, C. Lee, J. Hone, J. Shan and T. F. Heinz, *Phys Rev Lett*, 2010, **105**, 136805.
6. A. Splendiani, L. Sun, Y. Zhang, T. Li, J. Kim, C.-Y. Chim, G. Galli and F. Wang, *Nano Letters*, 2010, **10**, 1271-1275.
7. K. S. Novoselov, D. Jiang, F. Schedin, T. J. Booth, V. V. Khotkevich, S. V. Morozov and A. K. Geim, *P Natl Acad Sci USA*, 2005, **102**, 10451-10453.
8. B. Radisavljevic, A. Radenovic, J. Brivio, V. Giacometti and A. Kis, *Nat. Nanotechnol.*, 2011, **6**, 147-150.
9. N. Scheuschner, O. Ochedowski, A.-M. Kaulitz, R. Gillen, M. Schleberger and J. Maultzsch, *Phys Rev B*, 2014, **89**, 125406.
10. H. R. Gutiérrez, N. Perea-López, A. L. Elías, A. Berkdemir, B. Wang, R. Lv, F. López-Urías, V. H. Crespi, H. Terrones and M. Terrones, *Nano Letters*, 2012.
11. X. Wang, Y. Gong, G. Shi, W. L. Chow, K. Keyshar, G. Ye, R. Vajtai, J. Lou, Z. Liu, E. Ringe, B. K. Tay and P. M. Ajayan, *ACS Nano*, 2014.
12. M. O'Brien, N. McEvoy, T. Hallam, H.-Y. Kim, N. C. Berner, D. Hanlon, K. Lee, J. N. Coleman and G. S. Duesberg, *Sci. Rep.*, 2014, **4**, 7374.
13. D. Hanlon, C. Backes, T. M. Higgins, M. Hughes, A. O'Neill, P. King, N. McEvoy, G. S. Duesberg, B. Mendoza Sanchez, H. Pettersson, V. Nicolosi and J. N. Coleman, *Chemistry of Materials*, 2014, **26**, 1751-1763.
14. N. Scheuschner, R. Gillen, M. Staiger and J. Maultzsch, *arXiv preprint arXiv:1503.08980*, 2015.
15. H. Terrones, E. Del Corro, S. Feng, J. Poumirol, D. Rhodes, D. Smirnov, N. Pradhan, Z. Lin, M. Nguyen and A. Elías, *Scientific reports*, 2014, **4**.
16. P. Tonndorf, R. Schmidt, P. Böttger, X. Zhang, J. Börner, A. Liebig, M. Albrecht, C. Kloc, O. Gordan, D. R. T. Zahn, S. Michaelis de Vasconcellos and R. Bratschitsch, *Opt. Express*, 2013, **21**, 4908-4916.
17. H. Li, Q. Zhang, C. C. R. Yap, B. K. Tay, T. H. T. Edwin, A. Olivier and D. Baillargeat, *Advanced Functional Materials*, 2012, **22**, 1385-1390.
18. A. Berkdemir, H. R. Gutiérrez, A. R. Botello-Méndez, N. Perea-López, A. L. Elías, C.-I. Chia, B. Wang, V. H. Crespi, F. López-Urías and J.-C. Charlier, *Scientific reports*, 2013, **3**, 1755.
19. J. Xia, X. Huang, L.-Z. Liu, M. Wang, L. Wang, B. Huang, D.-D. Zhu, J.-J. Li, C.-Z. Gu and X.-M. Meng, *Nanoscale*, 2014, **6**, 8949-8955.
20. S.-Y. Chen, C. Zheng, M. S. Fuhrer and J. Yan, *Nano letters*, 2015.
21. X. Zhang, W. Han, J. Wu, S. Milana, Y. Lu, Q. Li, A. Ferrari and P. Tan, *Phys Rev B*, 2013, **87**, 115413.
22. X. Lu, J. Lin, M. I. B. Utama, X. Gong, J. Zhang, Y. Zhao, S. T. Pantelides, J. Wang, Z. Dong, Z. Liu, W. Zhou and Q. Xiong, *Nano Letters*, 2014.